\pdfoutput=1
\RequirePackage{ifpdf}
\ifpdf 
\documentclass[pdftex]{sigma}
\else
\documentclass{sigma}
\fi

\numberwithin{equation}{section}

\def\be{\begin{equation}}
\def\ee{\end{equation}}

\begin{document}

\allowdisplaybreaks

\newcommand{\arXivNumber}{1701.03057}

\renewcommand{\thefootnote}{}

\renewcommand{\PaperNumber}{047}

\FirstPageHeading

\ShortArticleName{Check-Operators and Quantum Spectral Curves}

\ArticleName{Check-Operators and Quantum Spectral Curves\footnote{This paper is a~contribution to the Special Issue on Combinatorics of Moduli Spaces: Integrability, Cohomo\-logy, Quantisation, and Beyond. The full collection is available at \href{http://www.emis.de/journals/SIGMA/moduli-spaces-2016.html}{http://www.emis.de/journals/SIGMA/moduli-spaces-2016.html}}}

\Author{Andrei MIRONOV~$^{\dag^1\dag^2\dag^3\dag^4}$ and Alexei MOROZOV~$^{\dag^2\dag^3\dag^4}$}

\AuthorNameForHeading{A.~Mironov and A.~Morozov}

\Address{$^{\dag^1}$~Lebedev Physics Institute, Moscow, 119991, Russia}
\EmailDD{\href{mailto:mironov@lpi.ru}{mironov@lpi.ru}}

\Address{$^{\dag^2}$~ITEP, Moscow, 117218, Russia}
\EmailDD{\href{mailto:morozov@itep.ru}{morozov@itep.ru}}

\Address{$^{\dag^3}$~Institute for Information Transmission Problems, Moscow, 127994, Russia}

\Address{$^{\dag^4}$~National Research Nuclear University MEPhI, Moscow, 115409, Russia}

\ArticleDates{Received January 29, 2017, in f\/inal form June 19, 2017; Published online June 26, 2017}

\Abstract{We review the basic properties of ef\/fective actions of families of theories (i.e., the actions depending on additional non-perturbative {\it moduli} along with perturbative couplings), and their description in terms of operators (called check-operators), which act on the moduli space. It is this approach that led to constructing the (quantum) spectral curves and what is now nicknamed the EO/AMM topological recursion. We explain how the non-commutative algebra of check-operators is related to the modular kernels and how symplectic (special) geometry emerges from it in the classical (Seiberg--Witten) limit, where the quantum integrable structures turn into the well studied classical integrability. As time goes, these results turn applicable to more and more theories of physical importance, supporting the old idea that many universality classes of low-energy ef\/fective theories contain matrix model representatives.}


\Classification{14H70; 81R10; 81R12; 81T13}

\renewcommand{\thefootnote}{\arabic{footnote}}
\setcounter{footnote}{0}

\section{Introduction}

\looseness=-1 One of the main lessons that we learned from string theory is that instead of working with a~particular model (string or f\/ield), one should better consider families of similar models at once. This provides the most ef\/fective description of the problem, no matter has one to integrate over the space of theories (as in string theory) or not (as in the ordinary quantum f\/ield theory). This approach allows one to reveal various non-trivial structures underlying the family of theories that require involving a dynamics on the moduli space of theories. The most instructive examples of these structures are the algebras of constraints (Ward identities), that act on the partition functions of the theories \cite{Vir3,Vir1,Vir6,Vir5,Vir4,Vir7,Vir2} and various integrable structures, from the standard integrability \cite{intSW2,MMint1,intSW1,intSW5,MMint5,MMint4,MMint6,MMint2,MMint3,intSW3,intSW4} to the Whitham integrability \cite{RG2, RG1}.

\looseness=-1 Technically, one of the most ef\/fective tools turned out to be matrix models, working out these simple examples it was possible to develop a notion of check-operators. These operators act on the moduli space, and it turned out to be possible to mimic the action of various operators of the concrete theory by an action of check-operators \cite{check2, check1}. This framework allowed one later to realize many essential structures, from the topological recursion\cite{toprec1a,toprec1b,toprec2,toprec3} to dualities~\cite{GMM2}, from the wall-crossing formulas~\cite{GMM3} to knot theory \cite{GMM3,GMMhik, Hikami1,Hikami2,Hikami5,Hikami3,Hikami4}. Moreover, it turns out that this approach provides a simple description of some standard objects like modular kernels \cite{GMM2} and of some standard procedures like quantization of Seiberg--Witten integrabilities \cite{MMMso}, both the standard \cite{qintSW2a,qintSW2b, qintSW1} and Whitham \cite{RG2,RG1,MMMso} ones. As predicted \cite{UFN232a,UFN232c,UFN232b,UFN231a,UFN231b,UFN231d,UFN231c}, nowadays they proved important far beyond the matrix model context, where they were originally discovered. The same will def\/initely happen to the new insights of general value from matrix model theory, which we mostly restrict
to in the present review. While 25 years ago one mostly dealt with integrability, Virasoro constraints (or loop equations, or Ward identities) genus expansions and spectral curves, today the top issues are check-operators, their action on spaces of solutions to the loop equations, the quantum spectral curves and quantized Whitham f\/lows.

Returning to applications, these include, f\/irst of all, topological strings. Here one considers matrix models of a quite general form which are referred to as matrix model networks \cite{MMnw3,KP,MMnw2c,MMnw2a,MMnw2b, MMnw1}. These matrix model networks have many avatars: on one hand, they may serve as a tool to study (ref\/ined) topological strings \cite{topstr2,topstr8a,topstr8b,topstr8c,topstr4,topstr1,topstr6,topstr7,topstr5,topstr3} and Nekrasov functions \cite{Nek2,Nek1,Nek3}, on another hand, within the framework of quantum f\/ield theory, they describe supersymmetric quiver gauge theories of Seiberg--Witten type, and, at last, at the algebraic level their partition functions are associated \cite{AGTmamo1,AGTmamo3a,AGTmamo3b,AGTmamo2,AGTmamo4c,AGTmamo5,AGTmamo4b,AGTmamo4a} with the conformal blocks of Virasoro/W and Ding--Iohara--Miki algebras. Interrelations between these subject is nothing but the AGT correspondence \cite{AGT1,AGTmamo1,AGTmamo3a,AGTmamo3b,AGTmamo2,AGTmamo4c,AGT3,AGTmamo5,AGTmamo4b,AGTmamo4a,AGT2}.

One can illustrate these relations with the best studied example of two-parametric deformations of Seiberg--Witten (SW) systems:
\begin{table}[h!]\centering\footnotesize
\begin{tabular}{|@{\,}c@{\,}|@{\,}c@{\,}|@{\,}c@{\,}|@{\,}c@{\,}|}
\hline
gauge theory&integrable system& spectral curve&AGT dual\\
\hline
Nekrasov function ($\epsilon_1,\epsilon_2$)&quantum integrable systems & degenerate conformal &conformal matrix\\
$\downarrow\epsilon_2\to 0$ &+ Whitham f\/lows&block equation& model, KP hierarchy\\
\hline
quantum Seiberg--Witten system&quantum many-body&quantum spectral curve&spectral dual\\
$\downarrow\epsilon_1\to 0$ &integrable system, $\hbar=\epsilon_1$&= Schr\"odinger (Baxter)& to quantum\\
&& equation & integrable system\\
\hline
Seiberg--Witten system&classical f\/inite-dimensional&spectral curve&spectral dual\\
&integrable system&(+ Whitham f\/lows)& to classical\\
&&& integrable system\\
\hline
\end{tabular}
\end{table}

Below we consider only the simplest realization of check-operators and their applications within the ordinary matrix models, not the matrix model networks. We demonstrate how to construct check-operators, to obtain SW systems and to generate quantum spectral curves using them. We also describe a simple application of check-operators: a derivation of the modular kernel in two-dimensional conformal f\/ield theory.

This paper is a review of earlier results mostly described in \cite{check2,check1,AMM12,AMM11,GMM2}. In particular, the check-operators (see examples in Sections~\ref{section3.1} and~\ref{section3.4} below) were introduced in \cite{check2, check1}, where their properties were discussed. Among the check-operators, there is the main check-operator (see Section~\ref{section3.5}) with the crucial property~(\ref{mp}). In a dif\/ferent situation, this main check-operator was discussed in~\cite{GMM2}.

\section{Multiple solutions to the Virasoro constraints}

\subsection{Simplest example: the Hermitean matrix integral}

Thus, we start with the simplest example of the Hermitean matrix model
\begin{gather}\label{MMP}
Z=\int {\rm d}M \exp[\operatorname{Tr} V(M)],
\end{gather}
where we parameterize
\begin{gather*}
V(M)=\sum_{k=0}t_kM^k
\end{gather*}
so that there is a natural grading $[t_k]=k$ and ${\rm d}M$ is the Haar measure of integration over $N\times N$ Hermitean matrices normalized to the volume of the unitary group~$U(N)$. $t_k$ are here coef\/f\/icients in the potential $V(M)$ which can be treated either as a formal series or as a~polynomial of large enough degree (in the sense of the projective limit).

This integral satisf\/ies an inf\/inite set of Virasoro constraints (=~loop equations) \cite{Vir3,Vir1,Vir6,Vir5, Vir4,Vir7,Vir2}:
\begin{gather}\label{Vir}
L_nZ=0,\qquad n\ge -1,\\
L_n=\sum t_k{\partial\over\partial t_{k+n}}+\sum_{a+b=n}{\partial^2\over\partial t_a\partial t_b},\nonumber\\
{\partial Z\over\partial t_0}=NZ.\nonumber
\end{gather}
These operators $L_n$ form a Borel subalgebra of the Virasoro algebra.

\subsection{Solutions as formal series}

The matrix model partition function is def\/ined in (\ref{MMP}) by a formal integral, which is still to be def\/ined. Instead of this, we def\/ine it as any solution to the constraints (\ref{Vir}). We also need to f\/ix a class of functions where we look for these solutions. Let us consider power series solutions in all $t_k$, i.e., we treat the integral (\ref{MMP}) perturbatively with respect to the potential $V(M)$. Then, there are no solutions to (\ref{Vir}) at all! This is because the moments $\langle M^k\rangle$ all diverge. One has to regularize them.

The simplest way to do this is to consider the Gaussian integral, i.e., make the substitution $t_2\to t_2-\alpha$. Then,
\begin{gather*}
Z=\int {\rm d}M \exp\big[{-}\alpha\operatorname{Tr} M^2+\operatorname{Tr} V(M)\big]=c_0+c_1t_1+c_2^{(1)}t_1^2+c_2^{(2)}t_2+\cdots\\
\hphantom{Z}{}
=\sum_n\sum_{\Delta\colon |\Delta|=n}c_n^\Delta t_\Delta,
\end{gather*}
where $c$'s are some coef\/f\/icients of grading $n$, which are constructed from the moments
\begin{gather*}
\int {\rm d}M \exp\big[{-}\alpha\operatorname{Tr} M^2\big] M^k,
\end{gather*}
and $\Delta$ is the Young diagram with lengths $\delta_1\ge \delta_2\ge\cdots\ge \delta_k$, $t_\Delta=\prod\limits_{i=1}^k t_{\delta_i}$.
Note that
\begin{gather*}
c_i\sim \alpha^{-i/2},
\end{gather*}
which is evident by dimensional argument.

Parameter $\alpha$ is the simplest example of a non-perturbative modulus. It is treated dif\/ferently from the perturbative couplings $t_k$, but not independent of them. It can appear in denominators of the coef\/f\/icients, still $\frac{\partial Z(\alpha|t)}{\partial \alpha} = -\frac{\partial Z(\alpha|t)}{\partial t_2}$.

\subsection{More general (Dijkgraaf--Vafa) case}

Let us now consider a more general Dijkgraaf--Vafa (DV) case, when a few f\/irst coef\/f\/icients $t_k$ are shifted $t_k\longrightarrow T_k+t_k$ so that the partition function (\ref{MMP}) becomes \cite{DV1,DV2,DV3}
\begin{gather}\label{DV}
Z_{\rm DV}=\int {\rm d}M \exp [ \operatorname{Tr} W(M)+\operatorname{Tr} V(M) ],
\end{gather}
where $W(M)=\sum^p T_kM^k$, and the integral is treated as a power series in~$t_k$'s, but as a function of $T_k$'s. One also has to make the shift $t_k\longrightarrow T_k+t_k$ in the Virasoro constraints.

Then, the coef\/f\/icients are constructed from
\begin{gather*}
\int {\rm d}M \exp [ \operatorname{Tr} W(M) ]M^k,
\end{gather*}
i.e., combinations of $T_k$'s appear in denominators. One understands these integrals as integrals over properly chosen contours at f\/ixed $W(M)$. This means that the matrices may not be literally Hermitean, since their eigenvalues are not obligatory real.

\subsection{How many solutions?}

Now one can enumerate the solutions to (\ref{Vir}) with $p$ f\/irst $t_k$'s shifted \cite{AMM12, AMM11}. One can state that solutions are parameterized by an arbitrary function of $p-2$ variables $T_k$. The two variables are f\/ixed by the only linear constraints
\begin{gather*}
L_0Z=0,\qquad L_{-1}Z=0,
\end{gather*}
and all other constraints do not impose more restrictions, but serve as recurrence relations that allow one to evaluate all coef\/f\/icients $c$. Thus, for $p=2$ (the Gaussian case) there is a unique solution.

\subsection{Technical tools: loop equations}

One can rewrite the constraints (\ref{Vir}) by introducing a generating functions of the connected correlators (resolvents) in the model (\ref{DV}):
\begin{gather*}
\rho^{(1)}(z)=\left\langle \operatorname{Tr}{1\over z-M}\right\rangle =\sum_{k=0}^\infty{1\over z^{k+1}}\big\langle \operatorname{Tr} M^k\big\rangle ={1\over Z_{\rm DV}}\hat\nabla_z Z_{\rm DV}=\hat\nabla_z {\cal F},\\
\rho^{(2)}(z_1,z_2)=\left\langle \operatorname{Tr}{1\over z_1-M}\operatorname{Tr}{1\over z_2-M}\right\rangle_c=\hat\nabla_{z_1}\hat\nabla_{z_2} {\cal F},\\
\ldots,
\end{gather*}
where
\begin{gather*}
\hat\nabla_z=\sum_{k=0}^\infty{1\over z^{k+1}}{\partial\over\partial t_k},\qquad Z_{\rm DV}=\exp{\cal F}.
\end{gather*}
Then, the generating function for the constraints (\ref{Vir})
\begin{gather*}
 [T(z)Z_{\rm DV} ]_-=0,\qquad T(z)\equiv\sum_{n=-\infty}^{+\infty}{L_n\over z^{n+2}}
\end{gather*}
can be rewritten in the form of loop equation
\begin{gather}\label{loopeq}
\rho^{(1)}(z)^2+\hat\nabla_z\rho^{(1)}(z)+W'(z)\rho^{(1)}(z)+\underbrace{\big[W'(z)\rho^{(1)}(z)\big]_+}_{\text{polynomial of degree } p-2}+\underbrace{\big[V'(z)\rho^{(1)}(z)\big]_-}_{=0\text{ as }t_k\to 0}=0.
\end{gather}
Here the indices ``$+$'' and ``$-$'' denote the non-negative power and negative power parts of the expression accordingly.

\section{Check-operators}
\subsection{Check-operator: acting on the space of solutions}\label{section3.1}

At all $t_k=0$, the last term in (\ref{loopeq}) vanishes, while the forth one
\begin{gather*}
f_{p-2}(z)= \big[W'(z)\rho^{(1)}(z)\big]_+
\end{gather*}
can be realized by the action of an operator $\check R_z$ in variables $T_k$ which are moduli of solutions
\begin{gather*}
f_{p-2}(z)\equiv \check R_z{\cal F},\qquad \check R_z=-\sum_{a,b}(a+b+2)T_{a+b+2}\ z^a{\partial\over\partial T_b}.
\end{gather*}
This is the f\/irst example of a check-operator \cite{AMM12, AMM11}.

This operator is crucially important to make the loop equations closed: $z$-dependence of $\rho$ depends on the action of $\check R_z$. At the same time, it af\/fects the equation only ``a little'': the corresponding piece is a polynomial of f\/inite degree in $z$, while the function $\rho^{(1)}(z)$ is essentially non-polynomial (in fact, one would better consider $\rho^{(1)}(z)dz$ as a 1-dif\/ferential \cite{toprec1a,toprec1b,toprec2,toprec3}, however, we do not discuss this kind of subtleties in the short review).

\subsection{Classical spectral curve}

Now let us def\/ine the classical spectral curve that describes the family of solutions to the matrix model\footnote{We remind here again that by the matrix model we mean the set of constraints (\ref{Vir}).}. To this end, one has to make the genus expansion by rescaling the variables
\begin{gather*}
(t_k,T_k)\to \left({1\over g}t_k,{1\over g}T_k\right),\qquad Z=\exp\left({1\over g^2}{\cal F}\right),
\end{gather*}
and considering the free energy expansion
\begin{gather*}
{\cal F}=\sum_k g^{2k}{\cal F}_k.
\end{gather*}
The leading term (planar limit) of this genus expansion (which corresponds to neglecting the second term in (\ref{loopeq})) in the resolvent is
\begin{gather*}
\rho^{(1)}_0(z)={-W'(z)+y(z)\over 2}
\end{gather*}
at all $t_k=0$, where
\begin{gather*}
y(z)^2\equiv W'(z)^2-4f_{p-2}(z)
\end{gather*}
determines the classical spectral curve. Generically, it is a hyperelliptic Riemann surface of genus $p-2$. Note that this resolvent is def\/ined as the generating function of correlators that are not just power series but functions of $T_k$'s.

Thus, the role of the check operator $\check R_z$ is exactly to provide the spectral curve. Loop equations then build $\rho(z)$ from this curve by a specially devised canonical procedure, known as the AMM/EO topological recursion \cite{toprec1a,toprec1b,toprec2,toprec3}. Remarkably, as we demonstrate below, these same check operators describe not only the classical spectral curve, but also its quantization.

\subsection{Examples}\label{Ex}

Let us consider a couple of simplest examples.
\begin{itemize}\itemsep=0pt
\item Gaussian case: $p=2$, $f_0(z)={\rm const}$, $y^2=z^2-{\rm const}$. This leads to the notorious semi-circle distribution \cite{Dyson}, and the Riemann surface has genus 0, it is just sphere.
\item Cubic polynomial $W_3$ case: $p=3$, $f_1(z)$ is a linear function, the spectral curve is a torus, the space of solutions is described by a function of one variable.
\end{itemize}

Let us discuss a meaning of this last example \cite{Mir}. In this case $p=3$, and, in accordance with general theory, the solutions to the Virasoro constraints (\ref{Vir}) are f\/ixed by a choice of an arbitrary function of one variable. What does this mean in terms of matrix integral?

For the eigenvalue matrix models it reduces to an $N$-fold integral over eigenvalues~$x_i$ of~$M$. Each of them, $\int {\rm d}x\, {\rm e}^{W_3(x)}$, depends on the choice of integration contours. For this cubic exponential, there are two independent contours (corresponding to two Airy functions), and any contour is an arbitrary linear combination of these two. Thus, the partition function is expanded into basic partition function with $N$ eigenvalues in the integrand parted into two groups (two possible contours) consisting of $N_1$ and $N_2$ eigenvalues, $N_1+N_2=N$. This describes the two-cut (torus) solution, and there is only one independent variable, say, the fraction $N_1/N_2$. This is why the solutions are parameterized by an arbitrary function of one variable. Increasing the degree of $W$ we get more and more independent integration contours and thus more and more moduli in the space of solutions.

\subsection{Summary of general properties}\label{section3.4}

Now we can formulate the general properties of solutions to the Virasoro constraints (\ref{Vir}) at f\/ixed $p$ and their moduli space \cite{check2, check1}.
\begin{itemize}\itemsep=0pt
\item[i)] Any solution is unambiguously labeled by an arbitrary function of $p-2$ $T$-variables. This function can be associated with the free energy at all $t_k=0$. We call it the {\it bare} free energy ${\cal F}^{(0)}(T)$.
\item[ii)] Solutions to the Virasoro constraints (or loop equations) are constructed from ${\cal F}^{(0)}(T)$ by an evolution operator $\hat U(T,t)$ that does not depend on~${\cal F}^{(0)}(T)$:
\begin{gather*}
Z(T,t)=\hat U(T,t){\rm e}^{{\cal F}^{(0)}(T)}.
\end{gather*}

\item[iii)] The evolution operator $\hat U(T,t)$ is understood here as a power series in $t_k$ with the coef\/f\/i\-cients which can be completely expressed in terms of the unique operator $\check R (x)$ with its non-local ``function''~$\check y$
\begin{gather*}
\check y\equiv \sqrt{W'(x)^2 -4\check R(x)},\qquad \check R (x)\equiv
-\sum_{a,b=0}(a+b+2)T_{a+b+2}x^a{\partial\over\partial T_b},
\end{gather*}
its derivatives and $W'(x)$. Here $\check y$ is def\/ined as a power series at large $x$, see \cite{check2, check1} for the details.
\end{itemize}

\subsection{Main check-operator}\label{section3.5}

These general properties have an immediate consequence: they allow one to introduce the notion of the main check-operator \cite{check2, check1}. Indeed, one can construct the resolvent from the free energy not only by the standard loop operator $\hat\nabla_z(t)$ acting on $t_k$'s, but also by a check-operator acting on the moduli $T_k$:
\begin{gather*}
\rho^{(1)}(z)=\hat\nabla_z(t){\cal F}=\check\nabla_z(T){\cal F}.
\end{gather*}
This check operator is called main, and it is for construction of this operator from the spectral curve ``bundle'' over the moduli space, that the AMM/EO recursion procedure was later devised \cite{toprec1a,toprec1b,toprec2,toprec3}. It follows from the previous subsection that {\it the main check-operator $\check\nabla_z$} is expressed through $y$, its derivatives and $W'(x)$. It is important to notice that $[\hat\nabla_{z_1}, \hat\nabla_{z_2}]=0$, but $[\check\nabla_{z_1},\check\nabla_{z_2}]\ne 0$. Hence, these operators are of dif\/ferent level of complexity, but the check-operator acts on a much smaller space. Unfortunately, many properties of the check-operator have not been well-studied yet, though some of them are already known.

\subsection{Main property}

It turns out that the main check-operator possesses a very crucial property \cite{check2, check1}:
\begin{gather}\label{mp}
\left[\oint_{A_i}{\rm d}z\check\nabla_z,\oint_{B_j}{\rm d}z\check\nabla_z\right]=\delta_{ij},
\end{gather}
where $A_i$ and $B_i$ are the $A$- and $B$-cycles over the classical spectral curve $y^2=W'^2(x)-4\check R(x){\cal F}^{(0)}$ and the statement has been checked at the vicinity of large~$x$, i.e., it requires a kind of analytic continuation to the whole spectral curve.

\section{Seiberg--Witten (SW) like solutions and integrable properties}

\subsection{DV/SW system}
The main check-operator property (\ref{mp}) immediately leads to the SW structure of the matrix models \cite{CMMV1,CMMV2,CM, DV1,DV2,DV3}. Indeed,
choose the basis of functions parameterizing the space of solutions to the Virasoro constraints (\ref{Vir})
to be eigenfunctions of the $A$-periods of the main check-operator: $\oint_{A_i}{\rm d}z\check\nabla_z Z_a=a_iZ_a$, i.e., $\oint_{A_i}{\rm d}z\check\nabla_z{\cal F}_a=\oint_{A_i}{\rm d}z\rho^{(1)}(z)=a_i$, then
\begin{gather*}
\oint_{B_i}\rho^{(1)}(z){\rm d}z={\partial {\cal F}_a\over\partial a_i}.
\end{gather*}
In the matrix model terms, the f\/illing numbers $N_i$'s that we discussed in Section~\ref{Ex} are asso\-cia\-ted with
\begin{gather*}
a_i=\oint_{A_i}\rho^{(1)}(z){\rm d}z.
\end{gather*}
Let us stress again that we consider all the objects being formal series w.r.t.\ variables~$t_k$'s, but functions of $T_k$'s, $N_i$'s, $a_i$'s.

\subsection{Integrable properties}

As usual, the matrix models have clear integrable properties:
\begin{itemize}\itemsep=0pt
\item $Z(N|t)$ (\ref{MMP}) is a $\tau$-function of the Toda chain (as a formal series) with $N$ playing role of the discrete time \cite{MMint1,MMint5,MMint4,MMint6,MMint2,MMint3}, while, in the DV case, it is a sum of~$Z_a$ (introduced in the previous subsection) which is this $\tau$-function \cite{MMZ}.
\item $Z_{\rm DV}(T_k,N_i)$ determines the SW system; hence, it satisf\/ies the Whitham hierarchy \cite{RG2, RG1} in the planar limit, and
$T_k$ are Whitham f\/lows \cite{CMMV1,CMMV2, CM}.
\item In the planar limit, $Z_{\rm DV}(T_k,N_i)$ as a function of $T_k$ and $N_i$ also satisf\/ies \cite{CMMV1,CMMV2} the WDVV equations \cite{WDVV2,WDVV3, WDVV1}, which is typical for $\tau$-functions of Whitham hierarchies, and for the SW systems \cite{genWDVV1a,genWDVV1b,genWDVV1c,genWDVV2}.
\item The Dijkgraaf--Vafa partition function $Z_{\rm DV}(T_k,N_i)$ as the SW system is also associated with a many-body integrable system, classical \cite{intSW2,intSW1,intSW5,intSW3,intSW4} in the planar limit or quantum \cite{MMMso,qintSW2a,qintSW2b,qintSW1}, maybe even with the Whitham f\/lows quantized \cite{MMMso} (see table in the Introduction). This quantization is realized by the check-operators.
\end{itemize}

\section{Quantum spectral curves}

The integrability of our matrix model allows one to def\/ine immediately the quantum spectral curve as an operator which cancels the Baker--Akhiezer function of the integrable system \cite{MMint1,MMint5,MMint4,MMint6, MMint2,MMint3,OG1, OG2}. Indeed, in our Toda chain case, the latter is def\/ined through the $\tau$-function (the matrix model partition function) as (in this subsection $V$ denotes the potential with {\it shifted} coef\/f\/icients, i.e., is a sum of $W+V$ in (\ref{DV}), and all the statements are treated in terms of formal series in~$V$)
\begin{gather*}
\Psi_{\rm BA}(z)={\rm e}^{V(z)/2}\Psi(z),
\end{gather*}
where
\begin{gather}\label{qsc1}
\Psi(z)={Z\left(t_k-{1\over kz^k}\right)\over Z(t)}={1\over Z(t)}z^N{\rm e}^{\int^z{\rm d}\xi\hat\nabla_\xi}Z(t)=\langle\det (z-M)\rangle.
\end{gather}
Here $\langle\cdots\rangle$ means the matrix model average. Since the Baker--Akhiezer function is proportional to the matrix model average of the determinant, one of the lessons is that this average also satisf\/ies the quantum spectral curve equation.

From the Virasoro constraints (\ref{Vir}), {\it the quantum spectral curve} looks like
\begin{gather}\label{qsc}
\left[\partial_z^2+V'(z)\partial_z+\check R_z\right]\Psi(z)=0
\end{gather}
and, then, the equation for the Baker--Akhiezer function is
\begin{gather*}
\left[\partial_z^2-{1\over 2}V''(z)+{1\over 4}V'(z)^2-{1\over 2}[\check R_zV(z)]+\check R_z\right]\Psi_{\rm BA}(z)=0.
\end{gather*}
In the classical (planar) limit, $\partial\log\Psi(z)=\rho^{(1)}_0(z)$ and equation (\ref{qsc}) turns into the classical spectral curve (planar loop equation):
\begin{gather*}
\rho^{(1)}_0(z)^2+V'(z)\rho^{(1)}_0(z)+\check R_z{\cal F}=0.
\end{gather*}
Note that, in integrable terms, $\check R_z$ contains the derivatives w.r.t.\ the Whitham times.

\section{Quantum curves from degenerate conformal blocks}

In the previous sections, we demonstrated what is the check-operator technique in the simplest example of matrix models. In the next two sections we illustrate it in a more involved example of two-dimensional conformal f\/ield theories \cite{CFT4,CFT1,CFT5,CFT3, CFT2}. The conformal block in this theory is also described by a matrix model, however, being a function, not just a formal series has more tricky global behaviour. This is one of the avatars of the AGT correspondence \cite{AGT1,AGT3,AGT2}, which implies that the conformal block can be described as a $\beta$-ensemble of the Dotsenko--Fateev type \cite{AGTmamo1,DF,AGTmamo3a,AGTmamo3b,AGTmamo2,AGTmamo4c,AGTmamo5,AGTmamo4b,AGTmamo4a}.

\subsection{AGT and degenerate conformal blocks: quantum spectral curve}

{\bf Conformal block.} The $n$-point conformal block $G(x_k,\Delta;\Delta_i,c)$ \cite{CFT4,CFT1,CFT5,CFT3, CFT2} depends on the external conformal dimensions~$\Delta_i$, on the internal dimension $\Delta$, on the central charge $c$ and on $n-3$ double ratios $x_k$ of points. These variables are most conveniently parameterized (in particular, from the point of view of the AGT correspondence) as $\Delta=(Q-\alpha)\alpha$, $c=1+6Q^2$, $Q=b-1/b$, the primary f\/ields can be written in terms of the free f\/ield $\phi(z)$ as $V_\alpha(z)= {:}{\rm e}^{i\alpha\phi(z)}{:}$ and ${:}\ldots{:}$ denotes the normal ordering.

{\bf Degenerate conformal block and the spectral curve.}
Let us suppose that one of the f\/ields in the conformal block is degenerate at a level~$L$, which means it is simultaneously a~primary f\/ield and a level $L$ descendant. Then, the corresponding conformal block satisf\/ies an equation of order~$L$ \cite{CFT4,CFT1,CFT5,CFT3, CFT2}. For instance,
$(b^2L_{-1}^2-L_{-2})V_{1/2b}(z)$ is a primary f\/ield, i.e., $V_{1/2b}(z)$ is degenerate at the second level. Then, the equation for the 5-point block with the degenerate f\/ield at $z$:
\begin{gather}
\bigg[ b^2z(z-1)\partial_z^2+(2z-1)\partial_z-\underbrace{{q(q\!-\!1)\over z\!-\!q}\partial_q\!+\text{rational function of }q}_{\text{check-operator}}\!\bigg]G_5(z|0,q,1,\infty)=0,\!\!\!\!\label{dcb}
\end{gather}
where $q$ is the double ratio of four other points and we placed three points at $0$, $1$ and $\infty$. This is the quantum spectral curve, while $q$ is a counterpart of $T_k$.

{\bf Comment on Toda quantum spectral curve.} In the limit when all $\Delta_i\to\infty$, this equation is reduced to the non-stationary Schr\"odinger ${\rm SU}(2)$ periodic Toda chain equation
\begin{gather*}
\left(\partial_z^2-2\Lambda^2\cosh z+{1\over 4}{\partial\over\partial\Lambda}\right)G_5^{\rm Toda}=0,
\end{gather*}
where $\Lambda$ is the limit of a properly rescaled variable $q$. This is the quantum spectral curve for the ${\rm SU}(2)$ periodic Toda chain, while {\it $\log\Lambda$ is known to play the role of the first Whitham time in the Seiberg--Witten theory}.

\subsection{Conformal matrix model}

Now let us note that the quantum spectral curve (\ref{dcb}) is the curve for a matrix model, namely for the {\it conformal matrix model} \cite{confMM4,confMM6,confMM5,AGTmamo1,AGTmamo3a,AGTmamo3b,AGTmamo2,confMM3,confMM1,AGTmamo4c,AGTmamo5,AGTmamo4b,AGTmamo4a, confMM2}:
\begin{gather}\label{cmm}
G_4(0,q,1,\infty)=q^{2\alpha_1\alpha_2}(1-q)^{2\alpha_2\alpha_3}\int\prod_i {\rm d}u_i \Delta^{2b^2}(u)u_i^{2b\alpha_1}(1-u_i)^{2b\alpha_3}(q-u_i)^{2b\alpha_2},
\end{gather}
where $\Delta(u)$ is the Van-der-Monde determinant and the integrals over $u_i$'s part into groups: there are two integration contours, $[0,q]$ and $[0,1]$. Then, $\alpha$, $\alpha_4$ are related to the number of these contours:
\begin{itemize}\itemsep=0pt
\item there are $N_1$ contours $[0,q]$ with
\begin{gather*}
bN_1=\alpha-\alpha_1-\alpha_2,
\end{gather*}
\item there are $N_2$ contours $[0,1]$ with
\begin{gather*}
bN_2=Q-\alpha-\alpha_3-\alpha_4.
\end{gather*}
\end{itemize}
$N_1$ and $N_2$ are associated with the Dijkgraaf--Vafa $N_i$ (see Section~\ref{Ex}). Since the $\beta$-en\-semb\-le~(\ref{cmm}) can be presented in the form
\begin{gather*}
G_4=\left\langle V_{\alpha_1}(0)V_{\alpha_2}(q)V_{\alpha_3}(1)V_{\alpha_4}(\infty)\left(\int_0^q V_b(u){\rm d}u\right)^{N_1}\left(\int_0^1 V_b(u){\rm d}u\right)^{N_2}\right\rangle_{\rm CFT},
\end{gather*}
where $\langle \cdots \rangle_{\rm CFT}$ denotes averaging in the free f\/ield theory and $\int V_b(u){\rm d}u$ is a screening charge, it is nothing but the four-point conformal block \cite{AGTmamo1,AGTmamo3a,AGTmamo3b,AGTmamo2,AGTmamo4c,AGTmamo5,AGTmamo4b,AGTmamo4a}. At the same time, the degenerate f\/ive-point conformal block $G_5=\langle V_{1/2b}(z)\cdots\rangle_{\rm CFT}$. Since $\langle V_{1/2b}(z)V_b(u)\rangle_{\rm CFT}=z-u$, one immediately obtains $G_5=\langle \det (z-u_i)\rangle $. From (\ref{qsc1}) one knows that the matrix model average of the determinant satisf\/ies the equation for the quantum spectral curve. Hence, the equation~(\ref{dcb}) for $G_5$ is {\it exactly the quantum spectral curve} for the Dotsenko--Fateev (or conformal) matrix model~(\ref{cmm}).

\section{Modular kernels in conformal f\/ield theory}

Now we are ready to use the developed technique to derive the modular kernel in conformal theory.

\subsection{Modular kernel for 4-point conformal block}

The modular kernel is def\/ined for the modular transformation $S\colon x\to 1-x$ by the formula
\begin{gather*}
G_4(x,a;a_i,b)=\int {\rm d}a' K(a,a';a_i,b)G_4(1-x,a';a_i,b),
\end{gather*}
and we use the notation $a_i=\alpha_i-Q/2$.

{\bf Explicit expression for modular kernel.} The explicit expression for the four-point conformal block was obtained by a tedious work in \cite{PT1,PT2} and has the form
\begin{gather*}
K(a,a';a_i,b)=4\sinh (2\pi a'/b )\sinh (2\pi ba' ) {S_b(u_1)S_b(u_2)\over S_b(v_1)S_b(v_2)}\int {\rm d}x \prod_{i=1}^4{S_b(x-\xi_i)\over S_b(x-\zeta_i)},
\end{gather*}
where $S_b(x)$ is the double sine function \cite{sfm1,sfm2,dsin2a,dsin2b,dsin2c, dsin1}, $u_i$, $v_i$, $\xi_i$, $\zeta_i$ are linear functions of $a_i$, $b$, $a$ and $a'$, and the choice of integration contours is quite tricky~\cite{PT1,PT2}.

{\bf Representation of $\boldsymbol{G(x,a;a_i,b)}$ as a $\boldsymbol{\beta}$-ensemble with $\boldsymbol{\beta=b^2}$.} One can also calculate the modular kernel from the matrix model representation of the conformal block~(\ref{cmm}) perturbatively in the genus expansion term by term \cite{GMM1,N}, the result being quite surprising: the modular kernel in all orders of the expansion is the Fourier kernel:
\begin{gather*}
K(a,a';a_i,b)={\rm e}^{2\pi iaa'}.
\end{gather*}
This results seems to contradict to the result of \cite{PT1,PT2}, and we now explain the reason for the dif\/ference and derive the result of \cite{PT1,PT2} in a simple way in a simpler case of the one-point conformal block on torus.

\subsection{1-point toric conformal block}

We consider the one-point toric conformal block, which has the following series expansion
\begin{gather*}
G(\tau,a;\mu)=1+q\left({\Delta_{\rm ext}(1-\Delta_{\rm ext})\over 2\Delta}+1\right)+O\big(q^2\big)
\end{gather*}
with $\Delta_{\rm ext}=\mu (Q-\mu)$, and $q=\exp i\pi\tau$, $\tau$ being the torus modular parameter. In terms of the AGT dual gauge theory, $\mu$ is the adjoint hypermultiplet mass. The modular transformation of the conformal block now is given by the modular transformation of torus:
\begin{gather*}
G(\tau,a;\mu)=\int {\rm d}a' K(a,a';\mu)G\big({-}\tau^{-1},a';\mu\big).
\end{gather*}

{\bf Explicit expression for modular kernel.} This time the explicit expression for the modular kernel due to~\cite{PTt} is
\begin{gather}\label{mkt}
K(a,a';\mu)\sim\int {\rm d}\xi {S_b(\xi+\mu/2-a')S_b(\xi+\mu/2+a')\over S_b(\xi+Q-\mu/2-a')S_b(\xi+Q-\mu/2+a')}{\rm e}^{4\pi ia\xi}.
\end{gather}

{\bf Modular kernel from $\boldsymbol{\beta}$-ensemble.} One can again get the modular kernel from the $\beta$-ensemble realization of the conformal block~\cite{GMM1,N}. In this case, the essential point is that the conformal block dif\/fers from the partition function of the $\beta$-ensemble by a normalization factor~\cite{GMM2}
\begin{gather*}
G(\tau,a;\mu)={1\over N(a)}Z(\tau,a;\mu),\qquad N(a)={\Gamma_b(2a+\mu)\Gamma_b(2a+Q-\mu)\over \Gamma_b(2a),
\Gamma_b(2a+Q)}
\end{gather*}
where $\Gamma_b(x)$ is the Barnes double gamma function \cite{BG1,BG2,BG3,sfm1,sfm2}. The partition function turns out again to be transformed in the genus expansion by the pure Fourier transform
\begin{gather*}
Z(\tau,a;\mu)=\int {\rm d}a' {\rm e}^{2\pi iaa'}Z(-\tau^{-1},a';\mu),
\end{gather*}
i.e., the modular kernel appeared to be purely exponential. We now see why this is not quite the case and explain how to correct the calculation.

\subsection{An archetypical example}

An archetypical example of duality is provided by the pair of operators constructed from the coordinate and momentum, $\hat A = {\rm e}^{i\hat {\cal P}}$ and $\hat B = {\rm e}^{i\hat {\cal Q}}$, with the commutation relation
\begin{gather}
\hat A\hat B = {\rm e}^{i\hbar}\hat B\hat A.\label{ABPQ}
\end{gather}
Then, their eigenfunctions are related by the Fourier transform in the eigenvalue space:
\begin{gather*} \hat A Z_a({\cal Q}) = {\rm e}^{ia}Z_a({\cal Q}), \qquad \hat B \tilde Z_{a'}({\cal Q})={\rm e}^{ia'}\tilde Z_{a'}({\cal Q})
\quad \stackrel{(\ref{ABPQ})}{\Longrightarrow} \quad Z_a({\cal Q}) = \int {\rm e}^{\frac{iaa'}{\hbar}}\tilde Z_{a'}({\cal Q}){\rm d}a',
\end{gather*}
which can be easily checked by the direct calculation of the eigenfunctions:
\begin{gather*}
Z_a({\cal Q}) = {\rm e}^{\frac{ia {\cal Q}}{\hbar}}, \qquad \tilde Z_{a'}({\cal Q}) = \delta({\cal Q}-a').
\end{gather*}

{\bf Check-operators.} One, however, does not need to calculate the eigenfunctions in order to determine what is the transformation kernel. Instead, one can substitute the two operators by their representatives in the eigenvalue space, which reproduce the right commutation relations:
\begin{gather*}\label{check}
\check A = {\rm e}^{ia}, \qquad \check B = {\rm e}^{{\hbar}\frac{\partial}{\partial a}}.
\end{gather*}
Then the transformation kernel $M(a,a')={\rm e}^{\frac{iaa'}{\hbar}}$ is simply obtained from the equation
\begin{gather}\label{kernel}
\check A(a) M(a,a') = \check B(a')M(a,a').
\end{gather}

\subsection{Conformal block as an eigenfunction}

The conformal block turns out to be an eigenfunction of some operator ${\cal L}_A$:
\begin{gather*}
{\cal L}_AG=\lambda G,\qquad {\cal L}_BG=\Lambda(\partial_{\lambda})G,
\end{gather*}
which is constructed, similarly to the previous subsection, from the canonical pair of opera\-tors~\cite{GMM2}. Taking into account the matrix model ($\beta$-ensemble) representation of the conformal block, it is natural that this pair is given by periods of the main check-operator,~(\ref{mp}). Hence,

{\bf Claim.}
\begin{gather*}
{\cal L}_\gamma={\rm e}^{b\oint_\gamma {\rm d}z\check\nabla_z}.
\end{gather*}

Since $[{\cal L}_A,{\cal L}_B]=1$, one obtains that {\it $K(a,a';\mu)$ is the Fourier exponential}. This is what was obtained perturbatively~\cite{GMM1,N}, and it was a pretty tedious calculation!

{\bf Subtlety.} Now one has to ask why (\ref{mkt}) is not the exponential. The answer is hidden in the analytic properties of the partition function: the conformal theory is invariant with respect to the ref\/lection $a\to -a$, but there are two dif\/ferent main check-operators
\begin{gather*}
\oint_{A}{\rm d}z\, \check\nabla_z^{(+)} Z^{(+)}_{a}=a Z^{(+)}_{a},\qquad \oint_{A}{\rm d}z\,\check\nabla_z^{(-)} Z^{(-)}_{a}=-a Z^{(-)}_{a},
\end{gather*}
and two dif\/ferent branches of the $\beta$-ensemble partition function, i.e., $G$ is globally def\/ined but~$Z(a)$ is {\it not}! There are two branches at $a>0$ and $a<0$. Thus, one should naturally act with a sum of two exponentials of the two main check-operators and take into account the normalization factor~$N(a)$ that recalculate the action of $\check\nabla_z$ from the partition function to the conformal block:
\begin{gather*}
{\cal L}_\gamma=\left[{1\over N(a)}{\rm e}^{b\oint_\gamma {\rm d}z\check\nabla^+_z}N(a)+{1\over N(-a)}{\rm e}^{-b\oint_\gamma {\rm d}z\check\nabla^-_z}N(-a)
\right].
\end{gather*}

\subsection{Modular kernel for the torus conformal block}

Now we are ready to calculate the exact modular kernel \cite{GMM2}.
First of all, one can realize the periods of check-operators in the space of eigenvalues similarly to~(\ref{check}):
\begin{gather*}
\oint_A {\rm d}z\check\nabla_z^{\pm}\to \pm 2\pi ia,\qquad \oint_B {\rm d}z\check\nabla_z^{\pm}\to \pm {1\over 2}{\partial_a}.
\end{gather*}
Thus, one obtains
\begin{gather*}
{\cal L}_B={\Gamma (2ab)\Gamma (bQ+2ab)\over\Gamma (b\mu+2ab)\Gamma (b(Q-\mu)+2ab)}{\rm e}^{{b\over 2}\partial_a}+(a\to -a).
\end{gather*}
Since ${\cal L}_A'=\cos 2\pi b a$, we can f\/ind the modular kernel from the equations (\ref{kernel}), which becomes
\begin{gather*}
\frac{1}{2}\left(\frac{\sin 2\pi b(a-\mu/2)}{\sin 2\pi ba}{\rm e}^{-\frac{b}{2}\partial_a}+
\frac{\sin 2\pi b(a+\mu/2)}{\sin 2\pi ba}{\rm e}^{\frac{b}{2}\partial_a}\right)K(a,a')=\cos 2\pi ba' K(a,a').
\end{gather*}
At large~$a$, only one exponential survives giving the pure exponential kernel (see next corrections in \cite{N2a,N2b}). The solution of the full equation is immediately constructed~\cite{GMM2} and is given by
\begin{gather*}
K(a,a';\mu)=\int {\rm d}\xi\, C_1(\xi)C_2(a'){S_b(\xi+\mu/2-a')S_b(\xi+\mu/2+a')\over S_b(\xi+Q-\mu/2-a')S_b(\xi+Q-\mu/2+a')}{\rm e}^{4\pi ia\xi},
\end{gather*}
where $C_1(\xi)$ is an arbitrary periodic function with period $b$ and $C_2(a')$ is an arbitrary function. This result coincides with formula~(4.41) in \cite{PTt} at $C_1=C_2=1$. Further details can be found in~\cite{N2b, N2a}.

\section{Conclusion}

In this review, we introduced and explained the very important notion of check-operator: the operator that acts on the moduli space of theories (or vacua/solutions). We constructed the operator manifestly in the simplest example of the Hermitian matrix model and in a more involved example of the two-dimensional conformal f\/ield theory, and demonstrated its use by deriving the corresponding Seiberg--Witten structures and the quantum spectral curves. We also illustrated the usefulness of the concept by a simple evaluation of the kernel of modular transformation of the conformal blocks done in terms of the check-operators. The calculation used the wonderful relation (\ref{mp}), which provides the impressive example of the properties and the relevance of check-operators for the quantization theory.

\subsection*{Acknowledgements}

This work was performed at the Institute for Information Transmission Problems with the f\/inancial support of the Russian Science Foundation (Grant No.14-50-00150).

\pdfbookmark[1]{References}{ref}
\LastPageEnding


\begin{thebibliography}{99}
\footnotesize\itemsep=0pt

\bibitem{topstr2}
Aganagic M., Klemm A., Mari\~no M., Vafa C., The topological vertex,
 \href{https://doi.org/10.1007/s00220-004-1162-z}{\textit{Comm. Math. Phys.}} \textbf{254} (2005), 425--478,
 \href{https://arxiv.org/abs/hep-th/0305132}{hep-th/0305132}.

\bibitem{AGT1}
Alday L.F., Gaiotto D., Tachikawa Y., Liouville correlation functions from
 four-dimensional gauge theories, \href{https://doi.org/10.1007/s11005-010-0369-5}{\textit{Lett. Math. Phys.}} \textbf{91}
 (2010), 167--197, \href{https://arxiv.org/abs/0906.3219}{arXiv:0906.3219}.

\bibitem{check2}
Alexandrov A., Mironov A., Morozov A., Solving {V}irasoro constraints in matrix
 models, \href{https://doi.org/10.1002/prop.200410212}{\textit{Fortschr. Phys.}} \textbf{53} (2005), 512--521,
 \href{https://arxiv.org/abs/hep-th/0412205}{hep-th/0412205}.

\bibitem{check1}
Alexandrov A., Mironov A., Morozov A., Unif\/ied description of correlators in
 non-{G}aussian phases of {H}ermitian matrix model, \href{https://doi.org/10.1142/S0217751X06029375}{\textit{Internat.~J.
 Modern Phys.~A}} \textbf{21} (2006), 2481--2517, \href{https://arxiv.org/abs/hep-th/0412099}{hep-th/0412099}.

\bibitem{toprec1a}
Alexandrov A., Mironov A., Morozov A., Instantons and merons in matrix models,
 \href{https://doi.org/10.1016/j.physd.2007.04.018}{\textit{Phys.~D}} \textbf{235} (2007), 126--167, \href{https://arxiv.org/abs/hep-th/0608228}{hep-th/0608228}.

\bibitem{toprec1b}
Alexandrov A., Mironov A., Morozov A., B{GWM} as second constituent of complex
 matrix model, \href{https://doi.org/10.1088/1126-6708/2009/12/053}{\textit{J.~High Energy Phys.}} \textbf{2009} (2009), no.~12,
 053, 49~pages, \href{https://arxiv.org/abs/0906.3305}{arXiv:0906.3305}.

\bibitem{AMM12}
Alexandrov A., Mironov A., Morozov A., Putrov P., Partition functions of matrix
 models as the f\/irst special functions of string theory. {II}.~{K}ontsevich
 model, \href{https://doi.org/10.1142/S0217751X09046278}{\textit{Internat.~J. Modern Phys.~A}} \textbf{24} (2009), 4939--4998,
 \href{https://arxiv.org/abs/0811.2825}{arXiv:0811.2825}.

\bibitem{AMM11}
Alexandrov A., Morozov A., Mironov A., Partition functions of matrix models:
 f\/irst special functions of string theory, \href{https://doi.org/10.1142/S0217751X04018245}{\textit{Internat.~J. Modern Phys.
 A}} \textbf{19} (2004), 4127--4163, \href{https://arxiv.org/abs/hep-th/0310113}{hep-th/0310113}.

\bibitem{CFT4}
Alvarez-Gaum\'e L., Random surfaces, statistical mechanics and string theory,
 \textit{Helv. Phys. Acta} \textbf{64} (1991), 359--526.

\bibitem{Vir3}
Ambj{\o}rn J., Makeenko Yu.M., Properties of loop equations for the {H}ermitian
 matrix model and for two-dimensional quantum gravity, \href{https://doi.org/10.1142/S0217732390001992}{\textit{Modern Phys.
 Lett.~A}} \textbf{5} (1990), 1753--1763.

\bibitem{topstr8a}
Awata H., Kanno H., Instanton counting, {M}acdonald function and the moduli
 space of {D}-branes, \href{https://doi.org/10.1088/1126-6708/2005/05/039}{\textit{J.~High Energy Phys.}} \textbf{2005} (2005),
 no.~5, 039, 26~pages, \href{https://arxiv.org/abs/hep-th/0502061}{hep-th/0502061}.

\bibitem{topstr8b}
Awata H., Kanno H., Ref\/ined {BPS} state counting from {N}ekrasov's formula and
 {M}acdonald functions, \href{https://doi.org/10.1142/S0217751X09043006}{\textit{Internat.~J. Modern Phys.~A}} \textbf{24}
 (2009), 2253--2306, \href{https://arxiv.org/abs/0805.0191}{arXiv:0805.0191}.

\bibitem{topstr8c}
Awata H., Kanno H., Changing the preferred direction of the ref\/ined topological
 vertex, \href{https://doi.org/10.1016/j.geomphys.2012.10.014}{\textit{J.~Geom. Phys.}} \textbf{64} (2013), 91--110,
 \href{https://arxiv.org/abs/0903.5383}{arXiv:0903.5383}.

\bibitem{MMnw3}
Awata H., Kanno H., Matsumoto T., Mironov A., Morozov A., Morozov A., Ohkubo
 Y., Zenkevich Y., Explicit examples of {DIM} constraints for network matrix
 models, \href{https://doi.org/10.1007/JHEP07(2016)103}{\textit{J.~High Energy Phys.}} \textbf{2016} (2016), no.~7, 103,
 67~pages, \href{https://arxiv.org/abs/1604.08366}{arXiv:1604.08366}.

\bibitem{confMM4}
Awata H., Matsuo Y., Odake S., Shiraishi J., Collective f\/ield theory,
 {C}alogero--{S}utherland model and ge\-ne\-ra\-lized matrix models, \href{https://doi.org/10.1016/0370-2693(95)00055-P}{\textit{Phys.
 Lett.~B}} \textbf{347} (1995), 49--55, \href{https://arxiv.org/abs/hep-th/9411053}{hep-th/9411053}.

\bibitem{confMM6}
Awata H., Matsuo Y., Odake S., Shiraishi J., Excited states of the
 {C}alogero--{S}utherland model and singular vectors of the {$W_N$} algebra,
 \href{https://doi.org/10.1016/0550-3213(95)00286-2}{\textit{Nuclear Phys.~B}} \textbf{449} (1995), 347--374,
 \href{https://arxiv.org/abs/hep-th/9503043}{hep-th/9503043}.

\bibitem{confMM5}
Awata H., Matsuo Y., Odake S., Shiraishi J., A Note on {C}alogero--{S}utherland
 model, {$W_n$} singular vectors and generalized matrix models,
 \textit{Soryushiron Kenkyu} \textbf{91} (1995), A69--A75,
 \href{https://arxiv.org/abs/hep-th/9503028}{hep-th/9503028}.

\bibitem{BG1}
Barnes E.W., The genesis of the double gamma functions, \href{https://doi.org/10.1112/plms/s1-31.1.358}{\textit{Proc. London
 Math. Soc.}} \textbf{S1-31} (1899), 358--381.

\bibitem{BG2}
Barnes E.W., The theory of the double gamma function, \href{https://doi.org/10.1098/rsta.1901.0006}{\textit{Philos. Trans.~R.
 Soc. Lond. Ser.~A}} \textbf{96} (1901), 265--387.

\bibitem{BG3}
Barnes E.W., On the theory of multiple gamma functions, \textit{Trans.
 Cambridge Philos. Soc.} \textbf{19} (1904), 374--425.

\bibitem{CFT1}
Belavin A.A., Polyakov A.M., Zamolodchikov A.B., Inf\/inite conformal symmetry in
 two-dimensional quantum f\/ield theory, \href{https://doi.org/10.1016/0550-3213(84)90052-X}{\textit{Nuclear Phys.~B}} \textbf{241}
 (1984), 333--380.

\bibitem{CMMV1}
Chekhov L., Marshakov A., Mironov A., Vasiliev D., D{V} and {WDVV},
 \href{https://doi.org/10.1016/S0370-2693(03)00543-4}{\textit{Phys. Lett.~B}} \textbf{562} (2003), 323--338,
 \href{https://arxiv.org/abs/hep-th/0301071}{hep-th/0301071}.

\bibitem{CMMV2}
Chekhov L., Marshakov A., Mironov A., Vasiliev D., Complex geometry of
 matrix models, \textit{Proc. Steklov Inst. Math.} \textbf{251} (2005),
 265--306, \href{https://arxiv.org/abs/hep-th/0506075}{hep-th/0506075}.

\bibitem{CM}
Chekhov L., Mironov A., Matrix models vs. {S}eiberg--{W}itten/{W}hitham
 theories, \href{https://doi.org/10.1016/S0370-2693(02)03163-5}{\textit{Phys. Lett.~B}} \textbf{552} (2003), 293--302,
 \href{https://arxiv.org/abs/hep-th/0209085}{hep-th/0209085}.

\bibitem{Vir1}
David F., Loop equations and nonperturbative ef\/fects in two-dimensional quantum
 gravity, \href{https://doi.org/10.1142/S0217732390001141}{\textit{Modern Phys. Lett.~A}} \textbf{5} (1990), 1019--1029.

\bibitem{CFT5}
Di~Francesco P., Mathieu P., S\'en\'echal D., Conformal f\/ield theory, \href{https://doi.org/10.1007/978-1-4612-2256-9}{\textit{Graduate
 Texts in Contemporary Physics}}, Springer-Verlag, New York, 1997.

\bibitem{DV1}
Dijkgraaf R., Vafa C., Matrix models, topological strings, and supersymmetric
 gauge theories, \href{https://doi.org/10.1016/S0550-3213(02)00766-6}{\textit{Nuclear Phys.~B}} \textbf{644} (2002), 3--20,
 \href{https://arxiv.org/abs/hep-th/0206255}{hep-th/0206255}.

\bibitem{DV2}
Dijkgraaf R., Vafa C., On geometry and matrix models, \href{https://doi.org/10.1016/S0550-3213(02)00764-2}{\textit{Nuclear Phys.~B}}
 \textbf{644} (2002), 21--39, \href{https://arxiv.org/abs/hep-th/0207106}{hep-th/0207106}.

\bibitem{DV3}
Dijkgraaf R., Vafa C., A perturbative window into non-perturbative physics,
 \href{https://arxiv.org/abs/hep-th/0208048}{hep-th/0208048}.

\bibitem{AGTmamo1}
Dijkgraaf R., Vafa C., Toda theories, matrix models, topological strings, and
 $N=2$ gauge systems, \href{https://arxiv.org/abs/0909.2453}{arXiv:0909.2453}.

\bibitem{Vir6}
Dijkgraaf R., Verlinde H., Verlinde E., Loop equations and {V}irasoro
 constraints in nonperturbative two-dimensional quantum gravity,
 \href{https://doi.org/10.1016/0550-3213(91)90199-8}{\textit{Nuclear Phys.~B}} \textbf{348} (1991), 435--456.

\bibitem{WDVV2}
Dijkgraaf R., Verlinde H., Verlinde E., Topological strings in {$d<1$},
 \href{https://doi.org/10.1016/0550-3213(91)90129-L}{\textit{Nuclear Phys.~B}} \textbf{352} (1991), 59--86.

\bibitem{intSW2}
Donagi R., Witten E., Supersymmetric {Y}ang--{M}ills theory and integrable
 systems, \href{https://doi.org/10.1016/0550-3213(95)00609-5}{\textit{Nuclear Phys.~B}} \textbf{460} (1996), 299--334,
 \href{https://arxiv.org/abs/hep-th/9510101}{hep-th/9510101}.

\bibitem{DF}
Dotsenko V.S., Fateev V.A., Conformal algebra and multipoint correlation
 functions in {$2$}{D} statistical models, \href{https://doi.org/10.1016/0550-3213(84)90269-4}{\textit{Nuclear Phys.~B}}
 \textbf{240} (1984), 312--348.

\bibitem{WDVV3}
Dubrovin B., Geometry of {$2$}{D} topological f\/ield theories, in Integrable
 Systems and Quantum Groups ({M}ontecatini {T}erme, 1993), \href{https://doi.org/10.1007/BFb0094793}{\textit{Lecture
 Notes in Math.}}, Vol.~1620, Springer, Berlin, 1996, 120--348,
 \href{https://arxiv.org/abs/hep-th/9407018}{hep-th/9407018}.

\bibitem{Dyson}
Dyson F.J., Statistical theory of the energy levels of complex systems.~{I},
 \href{https://doi.org/10.1063/1.1703773}{\textit{J.~Math. Phys.}} \textbf{3} (1962), 140--156.

\bibitem{topstr4}
Eguchi T., Kanno H., Topological strings and {N}ekrasov's formulas,
 \href{https://doi.org/10.1088/1126-6708/2003/12/006}{\textit{J.~High Energy Phys.}} \textbf{2003} (2003), no.~12, 006, 30~pages,
 \href{https://arxiv.org/abs/hep-th/0310235}{hep-th/0310235}.

\bibitem{AGTmamo3a}
Eguchi T., Maruyoshi K., Penner type matrix model and {S}eiberg--{W}itten
 theory, \href{https://doi.org/10.1007/JHEP02(2010)022}{\textit{J.~High Energy Phys.}} \textbf{2010} (2010), no.~2, 022,
 21~pages, \href{https://arxiv.org/abs/0911.4797}{arXiv:0911.4797}.

\bibitem{AGTmamo3b}
Eguchi T., Maruyoshi K., Seiberg--{W}itten theory, matrix model and {AGT}
 relation, \href{https://doi.org/10.1007/JHEP07(2010)081}{\textit{J.~High Energy Phys.}} \textbf{2010} (2010), no.~7, 081,
 18~pages, \href{https://arxiv.org/abs/1006.0828}{arXiv:1006.0828}.

\bibitem{toprec2}
Eynard B., Orantin N., Invariants of algebraic curves and topological
 expansion, \href{https://doi.org/10.4310/CNTP.2007.v1.n2.a4}{\textit{Commun. Number Theory Phys.}} \textbf{1} (2007), 347--452,
 \href{https://arxiv.org/abs/math-ph/0702045}{math-ph/0702045}.

\bibitem{Nek2}
Flume R., Poghossian R., An algorithm for the microscopic evaluation of the
 coef\/f\/icients of the {S}eiberg--{W}itten prepotential, \href{https://doi.org/10.1142/S0217751X03013685}{\textit{Internat.~J.
 Modern Phys.~A}} \textbf{18} (2003), 2541--2563, \href{https://arxiv.org/abs/hep-th/0208176}{hep-th/0208176}.

\bibitem{Vir5}
Fukuma M., Kawai H., Nakayama R., Continuum {S}chwinger--{D}yson equations and
 universal structures in two-dimensional quantum gravity, \href{https://doi.org/10.1142/S0217751X91000733}{\textit{Internat.~J.
 Modern Phys.~A}} \textbf{6} (1991), 1385--1406.

\bibitem{GMM1}
Galakhov D., Mironov A., Morozov A., S-duality as a {$\beta$}-deformed
 {F}ourier transform, \href{https://doi.org/10.1007/JHEP08(2012)067}{\textit{J.~High Energy Phys.}} \textbf{2012} (2012),
 no.~8, 067, 28~pages, \href{https://arxiv.org/abs/1205.4998}{arXiv:1205.4998}.

\bibitem{GMM2}
Galakhov D., Mironov A., Morozov A., {$S$}-duality and modular transformation
 as a non-perturbative deformation of the ordinary $pq$-duality,
 \href{https://doi.org/10.1007/JHEP06(2014)050}{\textit{J.~High Energy Phys.}} \textbf{2014} (2014), no.~6, 050, 24~pages,
 \href{https://arxiv.org/abs/1311.7069}{arXiv:1311.7069}.

\bibitem{GMM3}
Galakhov D., Mironov A., Morozov A., Wall crossing invariants: from quantum
 mechanics to knots, \href{https://doi.org/10.1134/S1063776115030206}{\textit{J.~Exp. Theor. Phys.}} \textbf{120} (2015),
 549--577, \href{https://arxiv.org/abs/1410.8482}{arXiv:1410.8482}.

\bibitem{GMMhik}
Galakhov D., Mironov A., Morozov A., {${\rm SU}(2)/{\rm SL}(2)$} knot
 invariants and {K}ontsevich--{S}oibelman monodromies, \href{https://doi.org/10.1134/S0040577916050056}{\textit{Theoret. and
 Math. Phys.}} \textbf{187} (2016), 678--694, \href{https://arxiv.org/abs/1510.05366}{arXiv:1510.05366}.

\bibitem{MMint1}
Gerasimov A., Marshakov A., Mironov A., Morozov A., Orlov A., Matrix models of
 two-dimensional gravity and {T}oda theory, \href{https://doi.org/10.1016/0550-3213(91)90482-D}{\textit{Nuclear Phys.~B}}
 \textbf{357} (1991), 565--618.

\bibitem{intSW1}
Gorsky A., Krichever I.M., Marshakov A., Mironov A., Morozov A., Integrability
 and {S}eiberg--{W}itten exact solution, \href{https://doi.org/10.1016/0370-2693(95)00723-X}{\textit{Phys. Lett.~B}} \textbf{355}
 (1995), 466--474, \href{https://arxiv.org/abs/hep-th/9505035}{hep-th/9505035}.

\bibitem{RG2}
Gorsky A., Marshakov A., Mironov A., Morozov A., R{G} equations from {W}hitham
 hierarchy, \href{https://doi.org/10.1016/S0550-3213(98)00315-0}{\textit{Nuclear Phys.~B}} \textbf{527} (1998), 690--716,
 \href{https://arxiv.org/abs/hep-th/9802004}{hep-th/9802004}.

\bibitem{intSW5}
Gorsky A., Mironov A., Integrable many-body systems and gauge theories,
 \href{https://arxiv.org/abs/hep-th/0011197}{hep-th/0011197}.

\bibitem{Hikami1}
Hikami K., Hyperbolic structure arising from a knot invariant,
 \href{https://doi.org/10.1142/S0217751X0100444X}{\textit{Internat.~J. Modern Phys.~A}} \textbf{16} (2001), 3309--3333,
 \href{https://arxiv.org/abs/math-ph/0105039}{math-ph/0105039}.

\bibitem{Hikami2}
Hikami K., Generalized volume conjecture and the {$A$}-polynomials: the
 {N}eumann--{Z}agier potential function as a classical limit of the partition
 function, \href{https://doi.org/10.1016/j.geomphys.2007.03.008}{\textit{J.~Geom. Phys.}} \textbf{57} (2007), 1895--1940,
 \href{https://arxiv.org/abs/math.QA/0604094}{math.QA/0604094}.

\bibitem{Hikami5}
Hikami K., Inoue R., Braiding operator via quantum cluster algebra,
 \href{https://doi.org/10.1088/1751-8113/47/47/474006}{\textit{J.~Phys.~A: Math. Theor.}} \textbf{47} (2014), 474006, 21~pages,
 \href{https://arxiv.org/abs/1404.2009}{arXiv:1404.2009}.

\bibitem{Hikami3}
Hikami K., Inoue R., Cluster algebra and complex volume of once-punctured torus
 bundles and 2-bridge links, \href{https://doi.org/10.1142/S0218216514500060}{\textit{J.~Knot Theory Ramifications}} \textbf{23}
 (2014), 1450006, 33~pages, \href{https://arxiv.org/abs/1212.6042}{arXiv:1212.6042}.

\bibitem{Hikami4}
Hikami K., Inoue R., Braids, complex volume and cluster algebras,
 \href{https://doi.org/10.2140/agt.2015.15.2175}{\textit{Algebr. Geom. Topol.}} \textbf{15} (2015), 2175--2194,
 \href{https://arxiv.org/abs/1304.4776}{arXiv:1304.4776}.

\bibitem{topstr1}
Iqbal A., All genus topological string amplitudes and 5-brane webs as {F}eynman
 diagrams, \href{https://arxiv.org/abs/hep-th/0207114}{hep-th/0207114}.

\bibitem{topstr6}
Iqbal A., Kashani-Poor A.K., The vertex on a strip, \href{https://doi.org/10.4310/ATMP.2006.v10.n3.a2}{\textit{Adv. Theor. Math.
 Phys.}} \textbf{10} (2006), 317--343, \mbox{\href{https://arxiv.org/abs/hep-th/0410174}{hep-th/0410174}}.

\bibitem{topstr7}
Iqbal A., Koz\c{c}az C., Vafa C., The ref\/ined topological vertex,
 \href{https://doi.org/10.1088/1126-6708/2009/10/069}{\textit{J.~High Energy Phys.}} \textbf{2009} (2009), no.~10, 069, 58~pages,
 \href{https://arxiv.org/abs/hep-th/0701156}{hep-th/0701156}.

\bibitem{topstr5}
Iqbal A., Vafa C., Nekrasov N., Okounkov A., Quantum foam and topological
 strings, \href{https://doi.org/10.1088/1126-6708/2008/04/011}{\textit{J.~High Energy Phys.}} \textbf{2008} (2008), no.~4, 011,
 47~pages, \href{https://arxiv.org/abs/hep-th/0312022}{hep-th/0312022}.

\bibitem{AGTmamo2}
Itoyama H., Maruyoshi K., Oota T., Notes on the quiver matrix model and 2d-4d
 conformal connection, \href{https://doi.org/10.1143/PTP.123.957}{\textit{Progr. Theoret. Phys.}} \textbf{123} (2010),
 957--987, \href{https://arxiv.org/abs/0911.4244}{arXiv:0911.4244}.

\bibitem{Vir4}
Itoyama H., Matsuo Y., Noncritical {V}irasoro algebra of the $d<1$ matrix model
 and the quantized string f\/ield, \href{https://doi.org/10.1016/0370-2693(91)90236-J}{\textit{Phys. Lett.~B}} \textbf{255} (1991),
 202--208.

\bibitem{sfm1}
Jimbo M., Miwa T., Quantum {KZ} equation with {$|q|=1$} and correlation
 functions of the {$XXZ$} model in the gapless regime, \href{https://doi.org/10.1088/0305-4470/29/12/005}{\textit{J.~Phys.~A:
 Math. Gen.}} \textbf{29} (1996), 2923--2958, \href{https://arxiv.org/abs/hep-th/9601135}{hep-th/9601135}.

\bibitem{sfm2}
Kharchev S., Lebedev D., Semenov-Tian-Shansky M., Unitary representations of
 {$U_q({\mathfrak{sl}}(2,{\mathbb R}))$}, the modular double and the
 multiparticle {$q$}-deformed {T}oda chains, \href{https://doi.org/10.1007/s002200100592}{\textit{Comm. Math. Phys.}}
 \textbf{225} (2002), 573--609, \href{https://arxiv.org/abs/hep-th/0102180}{hep-th/0102180}.

\bibitem{MMint5}
Kharchev S., Marshakov A., Mironov A., Morozov A., Generalized {K}ontsevich
 model versus {T}oda hierarchy and discrete matrix models, \href{https://doi.org/10.1016/0550-3213(93)90347-R}{\textit{Nuclear
 Phys.~B}} \textbf{397} (1993), 339--378, \href{https://arxiv.org/abs/hep-th/9203043}{hep-th/9203043}.

\bibitem{confMM3}
Kharchev S., Marshakov A., Mironov A., Morozov A., Pakuliak S., Conformal
 matrix models as an alternative to conventional multi-matrix models,
 \href{https://doi.org/10.1016/0550-3213(93)90595-G}{\textit{Nuclear Phys.~B}} \textbf{404} (1993), 717--750,
 \href{https://arxiv.org/abs/hep-th/9208044}{hep-th/9208044}.

\bibitem{MMint4}
Kharchev S., Marshakov A., Mironov A., Morozov A., Zabrodin A., Towards unif\/ied
 theory of {$2$}d gravity, \href{https://doi.org/10.1016/0550-3213(92)90521-C}{\textit{Nuclear Phys.~B}} \textbf{380} (1992),
 181--240, \href{https://arxiv.org/abs/hep-th/9201013}{hep-th/9201013}.

\bibitem{MMint6}
Kharchev S., Marshakov A., Mironov A., Morozov A., Zabrodin A., Unif\/ication
 of all string models with {$c<1$}, \href{https://doi.org/10.1016/0370-2693(92)91595-Z}{\textit{Phys. Lett.~B}} \textbf{275}
 (1992), 311--314, \href{https://arxiv.org/abs/hep-th/9111037}{hep-th/9111037}.

\bibitem{MMint2}
Kharchev S., Marshakov A., Mironov A., Orlov A., Zabrodin A., Matrix models
 among integrable theories: forced hierarchies and operator formalism,
 \href{https://doi.org/10.1016/0550-3213(91)90030-2}{\textit{Nuclear Phys.~B}} \textbf{366} (1991), 569--601.

\bibitem{MMint3}
Kharchev S., Mironov A., Integrable structures of unitary matrix models,
 \href{https://doi.org/10.1142/S0217751X92002179}{\textit{Internat.~J. Modern Phys.~A}} \textbf{7} (1992), 4803--4824.

\bibitem{KP}
Kimura T., Pestun V., Quiver W-algebras, \href{https://arxiv.org/abs/1512.08533}{arXiv:1512.08533}.

\bibitem{OG1}
Krichever I.M., Methods of algebraic geometry in the theory of non-linear
 equations, \href{https://doi.org/10.1070/RM1977v032n06ABEH003862}{\textit{Russian Math. Surveys}} \textbf{32} (1977), no.~6, 185--213.

\bibitem{RG1}
Krichever I.M., The {$\tau$}-function of the universal {W}hitham hierarchy,
 matrix models and topological f\/ield theories, \href{https://doi.org/10.1002/cpa.3160470403}{\textit{Comm. Pure Appl. Math.}}
 \textbf{47} (1994), 437--475, \href{https://arxiv.org/abs/hep-th/9205110}{hep-th/9205110}.

\bibitem{dsin2a}
Kurokawa N., Multiple sine functions and {S}elberg zeta functions,
 \href{https://doi.org/10.3792/pjaa.67.61}{\textit{Proc. Japan Acad. Ser.~A Math. Sci.}} \textbf{67} (1991), 61--64.

\bibitem{dsin2b}
Kurokawa N., Gamma factors and {P}lancherel measures, \href{https://doi.org/10.3792/pjaa.68.256}{\textit{Proc. Japan Acad.
 Ser.~A Math. Sci.}} \textbf{68} (1992), 256--260.

\bibitem{dsin2c}
Kurokawa N., Multiple zeta functions: an example, in Zeta Functions in Geometry
 ({T}okyo, 1990), \textit{Adv. Stud. Pure Math.}, Vol.~21, Kinokuniya, Tokyo,
 1992, 219--226.

\bibitem{Vir7}
Makeenko Yu., Marshakov A., Mironov A., Morozov A., Continuum versus discrete
 {V}irasoro in one-matrix models, \href{https://doi.org/10.1016/0550-3213(91)90379-C}{\textit{Nuclear Phys.~B}} \textbf{356}
 (1991), 574--628.

\bibitem{confMM1}
Marshakov A., Mironov A., Morozov A., Generalized matrix models as conformal
 f\/ield theories. {D}iscrete case, \href{https://doi.org/10.1016/0370-2693(91)90021-H}{\textit{Phys. Lett.~B}} \textbf{265} (1991),
 99--107.

\bibitem{genWDVV1a}
Marshakov A., Mironov A., Morozov A., W{DVV}-like equations in {${\mathcal
 N}=2$} {SUSY} {Y}ang--{M}ills theory, \href{https://doi.org/10.1016/S0370-2693(96)01231-2}{\textit{Phys. Lett.~B}} \textbf{389}
 (1996), 43--52, \href{https://arxiv.org/abs/hep-th/9607109}{hep-th/9607109}.

\bibitem{genWDVV1b}
Marshakov A., Mironov A., Morozov A., W{DVV} equations from algebra of forms,
 \href{https://doi.org/10.1142/S0217732397000807}{\textit{Modern Phys. Lett.~A}} \textbf{12} (1997), 773--787,
 \href{https://arxiv.org/abs/hep-th/9701014}{hep-th/9701014}.

\bibitem{genWDVV1c}
Marshakov A., Mironov A., Morozov A., More evidence for the {WDVV} equations in
 {${\mathcal N}=2$} {SUSY} {Y}ang--{M}ills theories, \href{https://doi.org/10.1142/S0217751X00000537}{\textit{Internat. J.
 Modern Phys.~A}} \textbf{15} (2000), 1157--1206, \href{https://arxiv.org/abs/hep-th/9701123}{hep-th/9701123}.

\bibitem{MMMso}
Marshakov A., Mironov A., Morosov A., On {AGT} relations with surface operator
 insertion and a stationary limit of beta-ensembles, \href{https://doi.org/10.1016/j.geomphys.2011.01.012}{\textit{J.~Geom. Phys.}}
 \textbf{61} (2011), 1203--1222, \href{https://arxiv.org/abs/1011.4491}{arXiv:1011.4491}.

\bibitem{intSW3}
Martinec E.J., Integrable structures in supersymmetric gauge and string theory,
 \href{https://doi.org/10.1016/0370-2693(95)01456-X}{\textit{Phys. Lett.~B}} \textbf{367} (1996), 91--96, \href{https://arxiv.org/abs/hep-th/9510204}{hep-th/9510204}.

\bibitem{intSW4}
Martinec E.J., Warner N.P., Integrable systems and supersymmetric gauge theory,
 \href{https://doi.org/10.1016/0550-3213(95)00588-9}{\textit{Nuclear Phys.~B}} \textbf{459} (1996), 97--112,
 \href{https://arxiv.org/abs/hep-th/9511052}{hep-th/9511052}.

\bibitem{UFN232a}
Mironov A., {$2$}{D} gravity and matrix models. {I}.~{$2$}{D} gravity,
 \href{https://doi.org/10.1142/S0217751X94001746}{\textit{Internat.~J. Modern Phys.~A}} \textbf{9} (1994), 4355--4405,
 \href{https://arxiv.org/abs/hep-th/9312212}{hep-th/9312212}.

\bibitem{UFN232c}
Mironov A., Quantum deformations of $\tau$-functions, bilinear identities and
 representation theory, \mbox{\href{https://arxiv.org/abs/hep-th/9409190}{hep-th/9409190}}.

\bibitem{genWDVV2}
Mironov A., W{DVV} equations and {S}eiberg--{W}itten theory, in Integrability:
 the {S}eiberg--{W}itten and {W}hitham Equations ({E}dinburgh, 1998), Editors
 H.W.~Braden, I.M.~Krichever, Gordon and Breach, Amsterdam, 2000, 103--123,
 \href{https://arxiv.org/abs/hep-th/9903088}{hep-th/9903088}.

\bibitem{UFN232b}
Mironov A., Matrix models of two-dimensional gravity, \textit{Phys. Part.
 Nuclei} \textbf{33} (2002), 1051--1145.

\bibitem{Mir}
Mironov A., Matrix models and matrix integrals, \href{https://doi.org/10.1007/s11232-006-0007-7}{\textit{Theoret. and Math.
 Phys.}} \textbf{146} (2006), 63--72, \href{https://arxiv.org/abs/hep-th/0506158}{hep-th/0506158}.

\bibitem{qintSW2a}
Mironov A., Morosov A., Nekrasov functions and exact {B}ohr--{S}ommerfeld
 integrals, \href{https://doi.org/10.1007/JHEP04(2010)040}{\textit{J.~High Energy Phys.}} \textbf{2010} (2010), no.~4, 040,
 15~pages, \href{https://arxiv.org/abs/0910.5670}{arXiv:0910.5670}.

\bibitem{AGTmamo4c}
Mironov A., Morosov A., Shakirov S., Brezin--{G}ross--{W}itten model as ``pure
 gauge'' limit of {S}elberg integrals, \href{https://doi.org/10.1007/JHEP03(2011)102}{\textit{J.~High Energy Phys.}}
 \textbf{2011} (2011), no.~3, 102, 25~pages, \href{https://arxiv.org/abs/1011.3481}{arXiv:1011.3481}.

\bibitem{Vir2}
Mironov A., Morozov A., On the origin of {V}irasoro constraints in matrix
 models: {L}agrangian approach, \href{https://doi.org/10.1016/0370-2693(90)91078-P}{\textit{Phys. Lett.~B}} \textbf{252} (1990),
 47--52.

\bibitem{qintSW2b}
Mironov A., Morozov A., Nekrasov functions from exact {B}ohr--{S}ommerfeld
 periods: the case of {${\rm SU}(N)$}, \href{https://doi.org/10.1088/1751-8113/43/19/195401}{\textit{J.~Phys.~A: Math. Theor.}}
 \textbf{43} (2010), 195401, 11~pages, \href{https://arxiv.org/abs/0911.2396}{arXiv:0911.2396}.

\bibitem{AGT3}
Mironov A., Morozov A., On {AGT} relation in the case of {${\rm U}(3)$},
 \href{https://doi.org/10.1016/j.nuclphysb.2009.09.011}{\textit{Nuclear Phys.~B}} \textbf{825} (2010), 1--37, \href{https://arxiv.org/abs/0908.2569}{arXiv:0908.2569}.

\bibitem{AGTmamo5}
Mironov A., Morozov A., Morozov A., Conformal blocks and generalized {S}elberg
 integrals, \href{https://doi.org/10.1016/j.nuclphysb.2010.10.016}{\textit{Nuclear Phys.~B}} \textbf{843} (2011), 534--557,
 \href{https://arxiv.org/abs/1003.5752}{arXiv:1003.5752}.

\bibitem{AGTmamo4b}
Mironov A., Morozov A., Shakirov S., Conformal blocks as {D}otsenko--{F}ateev
 integral discriminants, \href{https://doi.org/10.1142/S0217751X10049141}{\textit{Internat.~J. Modern Phys.~A}} \textbf{25}
 (2010), 3173--3207, \href{https://arxiv.org/abs/1001.0563}{arXiv:1001.0563}.

\bibitem{AGTmamo4a}
Mironov A., Morozov A., Shakirov S., Matrix model conjecture for exact {BS}
 periods and {N}ekrasov functions, \href{https://doi.org/10.1007/JHEP02(2010)030}{\textit{J.~High Energy Phys.}} \textbf{2010}
 (2010), no.~2, 030, 26~pages, \href{https://arxiv.org/abs/0911.5721}{arXiv:0911.5721}.

\bibitem{MMZ}
Mironov A., Morozov A., Zakirova Z., Comment on integrability in
 {D}ijkgraaf--{V}afa {$\beta$}-ensembles, \href{https://doi.org/10.1016/j.physletb.2012.04.036}{\textit{Phys. Lett.~B}} \textbf{711}
 (2012), 332--335, \href{https://arxiv.org/abs/1202.6029}{arXiv:1202.6029}.

\bibitem{MMnw2c}
Mironov A., Morozov A., Zenkevich Y., Ding--{I}ohara--{M}iki symmetry of
 network matrix models, \href{https://doi.org/10.1016/j.physletb.2016.09.033}{\textit{Phys. Lett.~B}} \textbf{762} (2016), 196--208,
 \href{https://arxiv.org/abs/1603.05467}{arXiv:1603.05467}.

\bibitem{MMnw2a}
Mironov A., Morozov A., Zenkevich Y., On elementary proof of {AGT} duality from
 six dimensions, \href{https://doi.org/10.1016/j.physletb.2016.03.006}{\textit{Phys. Lett.~B}} \textbf{756} (2016), 208--211,
 \href{https://arxiv.org/abs/1512.06701}{arXiv:1512.06701}.

\bibitem{MMnw2b}
Mironov A., Morozov A., Zenkevich Y., Spectral duality in elliptic systems,
 six-dimensional gauge theories and topological strings, \href{https://doi.org/10.1007/JHEP05(2016)121}{\textit{J. High
 Energy Phys.}} \textbf{2016} (2016), no.~5, 121, 44~pages,
 \href{https://arxiv.org/abs/1603.00304}{arXiv:1603.00304}.

\bibitem{confMM2}
Mironov A., Pakuliak S., On the continuum limit of the conformal matrix models,
 \href{https://doi.org/10.1007/BF01017146}{\textit{Theoret. and Math. Phys.}} \textbf{95} (1993), 604--625,
 \href{https://arxiv.org/abs/hep-th/9209100}{hep-th/9209100}.

\bibitem{CFT3}
Moore G., Seiberg N., Classical and quantum conformal f\/ield theory,
 \href{https://doi.org/10.1007/BF01238857}{\textit{Comm. Math. Phys.}} \textbf{123} (1989), 177--254.

\bibitem{UFN231a}
Morozov A., String theory: what is it?, \href{https://doi.org/10.1070/PU1992v035n08ABEH002255}{\textit{Phys. Usp.}} \textbf{35} (1992),
 671--714.

\bibitem{UFN231b}
Morozov A., Integrability and matrix models, \href{https://doi.org/10.1070/PU1994v037n01ABEH000001}{\textit{Phys. Usp.}} \textbf{37}
 (1994), 1--55, \href{https://arxiv.org/abs/hep-th/9303139}{hep-th/9303139}.
\bibitem{UFN231d}
Morozov A., Challenges of matrix models, \href{https://arxiv.org/abs/hep-th/0502010}{hep-th/0502010}.

\bibitem{UFN231c}
Morozov A., Matrix models as integrable systems, \href{https://arxiv.org/abs/hep-th/9502091}{hep-th/9502091}.

\bibitem{MMnw1}
Morozov A., Zenkevich Y., Decomposing {N}ekrasov decomposition, \href{https://doi.org/10.1007/JHEP02(2016)098}{\textit{J.~High
 Energy Phys.}} \textbf{2016} (2016), no.~2, 098, 44~pages,
 \href{https://arxiv.org/abs/1510.01896}{arXiv:1510.01896}.

\bibitem{Nek1}
Nekrasov N., Seiberg--{W}itten prepotential from instanton counting,
 \href{http://projecteuclid.org/euclid.atmp/1111510432}{\textit{Adv. Theor. Math. Phys.}} \textbf{7} (2003), 831--864,
 \href{https://arxiv.org/abs/hep-th/0206161}{hep-th/0206161}.

\bibitem{Nek3}
Nekrasov N., Okounkov A., Seiberg--{W}itten theory and random partitions,
 \href{https://arxiv.org/abs/hep-th/0306238}{hep-th/0306238}.

\bibitem{qintSW1}
Nekrasov N., Shatashvili S., Quantization of integrable systems and four
 dimensional gauge theories, in X{VI}th {I}nternational {C}ongress on
 {M}athematical {P}hysics, \href{https://doi.org/10.1142/9789814304634_0015}{World Sci. Publ.}, Hackensack, NJ, 2010, 265--289,
 \href{https://arxiv.org/abs/0908.4052}{arXiv:0908.4052}.

\bibitem{N}
Nemkov N., S-duality as {F}ourier transform for arbitrary $\epsilon_1$,
 $\epsilon_2$, \href{https://doi.org/10.1088/1751-8113/47/10/105401}{\textit{J.~Phys.~A: Math. Theor.}} \textbf{47} (2014), 105401,
 15~pages, \href{https://arxiv.org/abs/1307.0773}{arXiv:1307.0773}.

\bibitem{N2b}
Nemkov N., On modular transformations of toric conformal blocks,
 \href{https://doi.org/10.1007/JHEP10(2015)039}{\textit{J.~High Energy Phys.}} \textbf{2015} (2015), no.~10, 037, 26~pages,
 \href{https://arxiv.org/abs/1504.04360}{arXiv:1504.04360}.

\bibitem{N2a}
Nemkov N., Fusion transformations in {L}iouville theory, \href{https://doi.org/10.1134/S0040577916110040}{\textit{Theoret. and
 Math. Phys.}} \textbf{189} (2016), 1574--1591, \href{https://arxiv.org/abs/1409.3537}{arXiv:1409.3537}.

\bibitem{OG2}
Novikov S.P., A method for solving the periodic problem for the {K}d{V}
 equation and its generalizations, \href{https://doi.org/10.1216/RMJ-1978-8-1-83}{\textit{Rocky Mountain~J. Math.}} \textbf{8}
 (1978), 83--93.

\bibitem{topstr3}
Okounkov A., Reshetikhin N., Vafa C., Quantum {C}alabi--{Y}au and classical
 crystals, in The Unity of Mathe\-matics, \href{https://doi.org/10.1007/0-8176-4467-9_16}{\textit{Progr. Math.}}, Vol.~244,
 Birkh\"auser Boston, Boston, MA, 2006, 597--618, \href{https://arxiv.org/abs/hep-th/0309208}{hep-th/0309208}.

\bibitem{toprec3}
Orantin N., Symplectic invariants, {V}irasoro constraints and {G}ivental
 decomposition, \href{https://arxiv.org/abs/0808.0635}{arXiv:0808.0635}.

\bibitem{PT1}
Ponsot B., Teschner J., Liouville bootstrap via harmonic analysis on a
 noncompact quantum group, \mbox{\href{https://arxiv.org/abs/hep-th/9911110}{hep-th/9911110}}.

\bibitem{PT2}
Ponsot B., Teschner J., Clebsch--{G}ordan and {R}acah--{W}igner coef\/f\/icients
 for a continuous series of representations of {$\mathcal
 U_q(\mathfrak{sl}(2,{\mathbb R}))$}, \href{https://doi.org/10.1007/PL00005590}{\textit{Comm. Math. Phys.}} \textbf{224}
 (2001), 613--655, \href{https://arxiv.org/abs/math.QA/0007097}{math.QA/0007097}.

\bibitem{dsin1}
Shintani T., On a {K}ronecker limit formula for real quadratic f\/ields,
 \textit{J.~Fac. Sci. Univ. Tokyo Sect. IA Math.} \textbf{24} (1977),
 167--199.

\bibitem{PTt}
Teschner J., From {L}iouville theory to the quantum geometry of {R}iemann
 surfaces, \href{https://arxiv.org/abs/hep-th/0308031}{hep-th/0308031}.

\bibitem{WDVV1}
Witten E., On the structure of the topological phase of two-dimensional
 gravity, \href{https://doi.org/10.1016/0550-3213(90)90449-N}{\textit{Nuclear Phys.~B}} \textbf{340} (1990), 281--332.

\bibitem{AGT2}
Wyllard N., {$A_{N-1}$} conformal {T}oda f\/ield theory correlation functions
 from conformal {${\mathcal N}=2$} {${\rm SU}(N)$} quiver gauge theories,
 \href{https://doi.org/10.1088/1126-6708/2009/11/002}{\textit{J.~High Energy Phys.}} \textbf{2009} (2009), no.~11, 002, 22~pages,
 \href{https://arxiv.org/abs/0907.2189}{arXiv:0907.2189}.

\bibitem{CFT2}
Zamolodchikov A.B., Zamolodchikov A.B., Conformal f\/ield theory and critical
 phenomena in two-dimensional systems, \textit{Soviet Sci. Rev.~A~Phys.}
 \textbf{10} (1989), 269--433.

\end{thebibliography}
\end{document}